\documentclass[aps,prl,epsf,twocolumn]{revtex4}
\usepackage{graphicx,epsf}
\usepackage{dcolumn}
\usepackage{bm}

\begin{document}


\def\appConv{Appendix~A}
\def\appBeer{Appendix~B}
\def\appHValid{Appendix~C}

\def\ia3{$Ia\overline{3}$}
\def\mn2o3{{$\alpha$-Mn$_2$O$_3$}}
\def\HfO2{{HfO$_2$}}
\def\SiO2{{SiO$_2$}}
\def\Al2O3{{Al$_2$O$_3$}}
\def\k{{$\kappa$~}}
\def\ks{{$\kappa_s$~}}
\def\ksten{{$\tensor{\kappa}_s$~}}
\def\kinf{{$\kappa_\infty$~}}
\def\kion{{$\kappa_{ion}$~}}
\def\kionten{{$\tensor{\kappa}_{ion}$~}}
\def\kinften{{$\tensor{\kappa}_\infty$~}}
\def\cm1{{cm$^{-1}$}}
\def\zstar{{$Z^{\star}$~}}

\def\comment#1{{\large\textsl{#1}}}
\def\degree {{$^\circ$}}
\def\degrees{{$^\circ$}}
\def\eq#1{{Eq.~(\ref{eq:#1})}}
\def\fig#1{{Fig.~\ref{fig:#1}}}
\def\sec#1{{Sec.~\ref{sec:#1}}}
\def\inv{^{-1}}
\def\micron {\hbox{$\mu$m}}
\def\microns{\micron}
\def\Ref#1{{Ref.~\onlinecite{#1}}}  
\def\tab#1{{Table~\ref{tab:#1}}}
\def\tauOne{\tau^{(1)}}
\def\tVec{\hbox{\bf t}}
\def\thetaDet{\theta_{DET}}

\def\qvec{{\vec q}}
\def\pvec{{\vec p}}
\def\Avec{{\vec A}}
\def\qhat{{\hat q}}
\def\qperphat{{\hat q_\perp}}
\def\ekpq{{E_{\kvec+\qvec}}}
\def\ek{{E_{\kvec}}}
\def\Omegabar{{\bar\Omega}}
\def\omegabar{{\bar\omega}}
\def\omegap{{\omega_p}}
\def\kf{{k_F}}
\def\kappaf{{\kappa_F}}
\def\mone{{-1}}
\def\re{{\rm{Re\,}}}
\def\im{{\rm{Im\,}}}
\def\twopi{{2 \pi}}
\def\wpm{w_\pm}
\def\FWlindhard{Appendix~A}
\def\lindhardTrans{Appendix~B}
\def\ftr{{f^{tr}}}
\def\PN{Pines and Nozi{\`e}res}
\def\Bohm{B{\"o}hm}
\def\Nifosi{Nifos{\'\i}}
\def\prin{{\cal P}}

\def\imagOmegaSq{the Appendix}
\def\epsTensor{{\buildrel \leftrightarrow \over \epsilon}}
\def\chiTensor{{\buildrel \leftrightarrow \over \chi}}
\def\idenTensor{{\buildrel \leftrightarrow \over I}}
\def\epsTrans{{\epsilon^{(t)}}}
\def\epsLong{{\epsilon^{(\ell)}}}
\def\epsTransInv{{\epsilon^{(t)-1}}}
\def\epsLongInv{{\epsilon^{(\ell)-1}}}

\def\MvecA{{M^{(\vec A)}}}
\def\MdivA{{M^{(\nabla \cdot \vec A)}}}
\def\Mphi{{M^{(\phi)}}}
\def\backGrad{{\buildrel \leftarrow \over \nabla}}
\def\backMom{i \hbar \backGrad}

\def\half{{1/2}}
\def\minusHalf{{-1/2}}
\def\threeHalves{{3/2}}
\def\minusThreeHalves{{-3/2}}

\newenvironment{bulletList}{\begin{list}{$\bullet$}{}}{\end{list}}

\title{The magnetic structure of bixbyite $\alpha$-Mn$_2$O$_3$: a combined density functional theory DFT+U and neutron diffraction 
study}

\author{Eric Cockayne$^1$}
\author{Igor Levin$^1$}
\author{Hui Wu$^{1,2}$}
\author{Anna Llobet$^3$}

\affiliation{$^1$National Institute of Standards and
Technology, Gaithersburg, Maryland 20899 USA}

\affiliation{$^2$Department of Materials Science and Engineering,
University of Maryland, College Park, Maryland 20742 USA}

\affiliation{$^3$Lujan Neutron Scattering Center, Los Alamos National 
Laboratory,  Los Alamos, NM 87545 USA}
 
\date{\today}

\begin{abstract}

First principles density functional theory DFT+U calculations and experimental 
neutron diffraction structure analyses were used to determine 
the low-temperature crystallographic and magnetic structure of bixbyite \mn2o3.  
The energies of various magnetic arrangements, calculated from first principles, 
were fit to a cluster-expansion model using a Bayesian method that overcomes a 
problem of underfitting caused by the limited number of input magnetic configurations.     
The model was used to predict the lowest-energy magnetic states.  
Experimental determination of magnetic structure benefited from optimized sample synthesis, 
which produced crystallite sizes large enough to yield a clear splitting of peaks in the 
neutron powder diffraction patterns, thereby enabling magnetic-structure refinements under 
the correct orthorhombic symmetry.   The refinements employed group theory  to 
constrain magnetic models.  Computational and experimental analyses independently 
converged to similar ground states, with identical antiferromagnetic ordering along a 
principal magnetic axis and secondary ordering along a single orthogonal axis, 
differing only by a phase factor in the modulation patterns. 
The lowest-energy magnetic states are compromise solutions to 
frustrated antiferromagnetic interactions between certain corner-sharing [MnO$_6$] octahedra.



\end{abstract}


\maketitle
\thispagestyle{empty}

\section*{Introduction}
\label{sec:intro}

 Manganese is a multivalent element.  Each valence state has a 
characteristic oxide or oxides.  
Mn ion magnetism and electron correlations of Mn $d$ states 
make it challenging to accurately determine the electronic structure of 
Mn oxides using the density functional theory (DFT) approach. 
Franchini {\it et al.}\cite{Franchini07}
explored various manganese oxides using DFT and showed that
the electronic structure as well as the magnetic ground
state depend sensitively on the details of the DFT calculation,
such as the exchange-correlation functional used.

\begin{figure}[h]
\includegraphics[width=75mm]{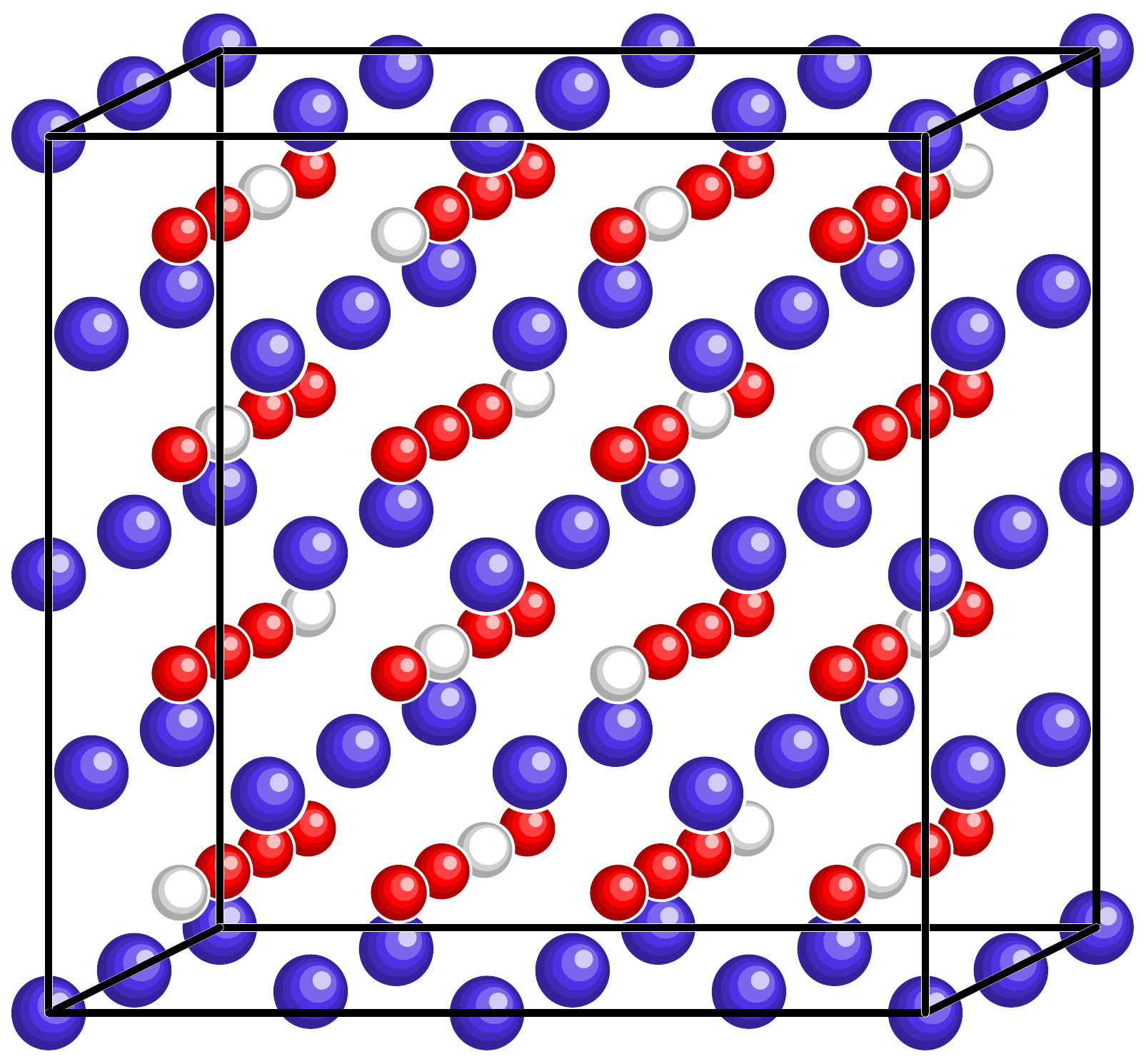}
\caption{Topology of bixbyite phase.  Mn in dark blue;
O in light red; unfilled tetrahedral interstitials indicated
in white}
\label{fig:bixby}
\end{figure}

\begin{figure}[h]
\includegraphics[width=75mm]{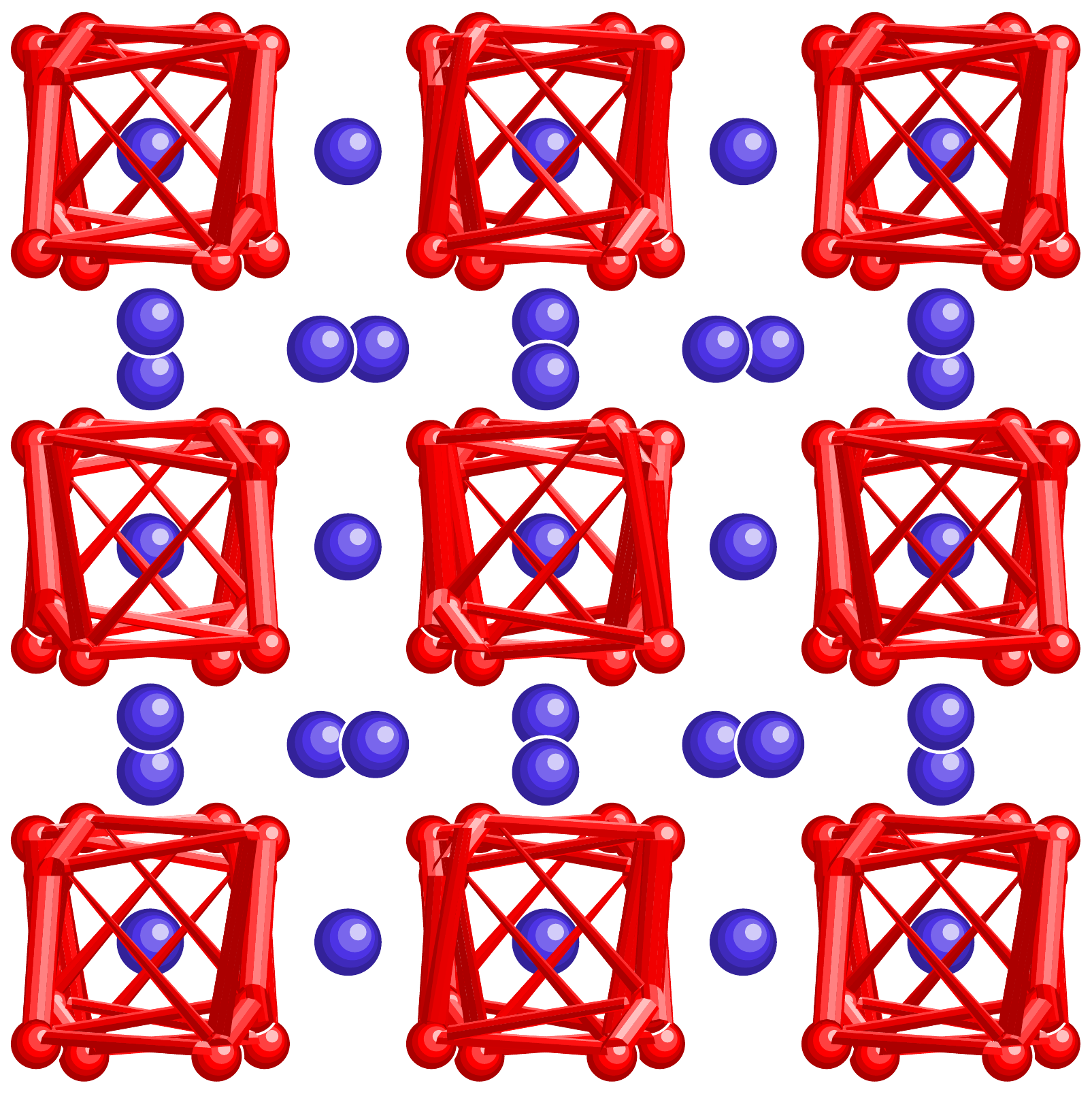}
\caption{Cubic \mn2o3 viewed along $b$ axis; regular octahedra shown.}
\label{fig:cubic}
\end{figure}

\begin{figure}[h]
\includegraphics[width=75mm]{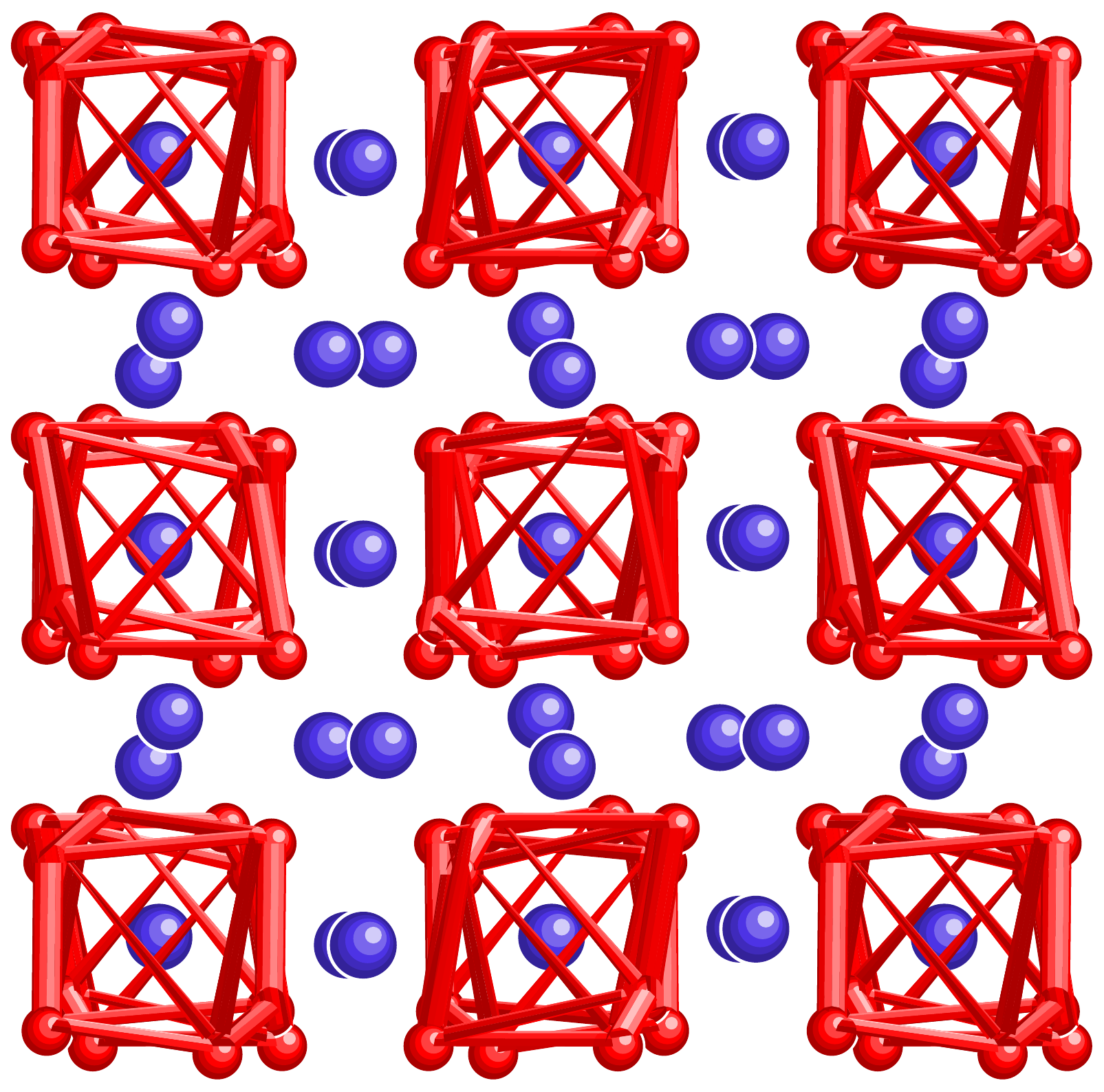}
\caption{Orthorhombic \mn2o3: octahedra shown in
\fig{cubic} have undergone Jahn-Teller distortion.}
\label{fig:ortho}
\end{figure}

 The $\alpha$-Mn$_2$O$_3$ phase, with the bixbyite structure,
is particularly challenging to model because of its complicated,
and not yet completely solved, magnetic structure.
In the bixbyite structure (\fig{bixby}), the Mn$^{3+}$ ions are octahedrally
coordinated, while the O ions have 4 Mn neighbors.
The bixbyite structure can be viewed as
a close-packed lattice of Mn with O ions filling 3/4 of
the tetrahedral interstitials in a pattern 
with $Ia\overline{3}$ symmetry.
Below about 300 K, $\alpha$-Mn$_2$O$_3$ transforms from cubic
to an orthorhombic structure with $Pbca$ symmetry\cite{LB75}.  
Experimentally, the lattice parameters ($a$,$b$,$c$) saturate at 
low temperatures to approximately $a$= 9.41~\AA, $b$ = 9.45~\AA, and
$c$ = 9.37~\AA~(Refs.~\onlinecite{Geller70,LB75,foot2}).
Geller\cite{Geller71} rationalized the low-temperature 
orthorhombic distortion of bixbyite as a consequence of a
Jahn-Teller instability of [MnO$_6$] octahedra toward  elongation
along any one of the three Cartesian axes.   In the cubic bixbyite phase,
24 of 32 Mn atoms exhibit distorted coordination whereas the remaining
8 Mn atoms, which occupy fixed-coordinate high-symmetry positions,
retain regular coordination environments.
The orthorhombic phase accommodates
Jahn-Teller distortion of the remaining 8 octahedra 
(see \fig{cubic}-\fig{ortho}).

 The magnetic structure of Mn$_2$O$_3$ has long been of
interest, but it is not completely solved.  
Computationally, Franchini {\it et al.} found a preference for either
antiferromagnetic or ferromagnetic ordering in \mn2o3, depending on the type
of DFT exchange-correlation functional used\cite{Franchini07}.
Experimentally, Regulski {\it et al.}\cite{Regulski04} found evidence
for various antiferromagnetic ordering transitions within the
orthorhombic phase, which occur without any apparent change in symmetry.

 Grant {\it et al.}\cite{Grant68} suggested that the magnetic ordering of
orthorhombic \mn2o3 can be predicted from the cubic 
$Ia\overline{3}$ symmetry because the orthorhombic distortion is small.
They proposed a non-collinear ordering model with magnetic moments on
the Mn 8($b$) sites (4($a$) and 4($b$) in $Pbca$) aligned with the body diagonals of
the pseudo-cubic cell and those on the Mn 24($d$) sites (8($c$) in $Pbca$) directed
perpendicular to one of the two orthogonal 2$_1$ screw axes passing through
each 24($d$) site.   However, Regulski {\it et al.} demonstrated that this model
is incompatible with the neutron powder diffraction data\cite{Regulski04}.  
They further identified an alternative, better-fitting, collinear model which 
featured antiferromagnetic ordering
on each of the five inequivalent Mn sublattices of the orthorhombic structure.
However, the orthorhombic distortion could not be resolved in the
diffraction patterns used by Regulski {\it et al} and, therefore, the atomic
positions had to be refined according to the high-temperature cubic 
$Ia\overline{3}$
structure.  No refinements of the nuclear and magnetic structures of the
magnetic phase under the correct orthorhombic symmetry have been reported.

 In this paper, we address deficiencies in both the computational
and experimental studies of the magnetic ground state of $\alpha$-Mn$_2$O$_3$.
Computationally, we use accurate parameterization of the exchange correlation and
onsite Hubbard parameters at the density functional theory  
DFT + U level in concurrence with a cluster-expansion model 
to investigate candidate ground state models until the correct DFT ground
state is established.  Experimentally, we use optimized conditions to synthesize
\mn2o3 with crystallite sizes large enough to yield visible
splitting of reflection peaks in the neutron diffraction patterns of
the orthorhombic phase, and  thus refine the magnetic ordering
within orthorhombic symmetry.  Both approaches give very similar
results for the magnetic ordering, suggesting that the ground state magnetic
structure of \mn2o3 is largely solved.

\section*{Computational Methods}

First principles density functional theory (DFT) calculations, as encoded in
the VASP software\cite{Kresse96,disclaim}, were used to calculate the 
relaxed configurations investigated here and their electronic 
structures.  The generalized gradient approximation (GGA) for the
exchange-correlation functional was used throughout, within the
``Perdew-Burke-Ernzerhof revised for solids" or ``PBEsol"
parameterization\cite{Perdew08}.

The PBEsol exchange-correlation functional has been found to 
give excellent results compared with experiment 
for the lattice parameters and bulk moduli of both metals
and nonmetals\cite{Csonka09}.  As is generally true for 
DFT, however, calculated band gaps are too small.  This error
can lead to qualitative errors for narrow bandgap 
materials and for materials with magnetic ions, both which
are true for \mn2o3.
We compensated for this error by including
onsite Coulomb terms (the ``GGA+U" approximation).
In a previous study of MnO$_2$ phases\cite{Cockayne12}, 
we found that the experimental volume and bandgap of 
$\beta$-MnO$_2$ could both be reproduced  using the
rotationally invariant DFT+U of 
Liechtenstein {\it et al.}\cite{Liech95} with an
onsite Coulomb potential $J$ = 2.8 eV and onsite exchange potential
$U$ = 1.2 eV for Mn $d$ electrons.
Remarkably, these same values used for the Mn$^{4+}$ ion 
of $\beta$-MnO$_2$ ion were found to be transferable to 
the Mn$^{3+}$ ion of \mn2o3, giving excellent agreement
with experiment, as shown below.
{\it Ex post facto} investigations of the effects of
varying the Mn $U$ and $J$ parameters, or adding $U$ or $J$ parameters
for oxygen, gave little, if any, improvement.

 Sufficient convergence in total energies and lattice parameters
was achieved with a plane wave cutoff energy of 500 eV and a 
2$\times$2$\times$2 Monkhorst-Pack grid of k-points.  A 
8$\times$8$\times$8 Monkhorst-Pack grid was used for the
density of states calculation.  The magnetic ordering of each 
of the 32 Mn atoms in the unit cell could be set 
either ``up" or ``down" as desired.  Spin-orbit coupling was 
neglected.  Only collinear magnetic structures were computed using VASP.

 The aim of the computational work was to find the DFT ground state
magnetic ordering of \mn2o3.  The cluster-expansion concept,
as developed for interatomic alloys\cite{Connolly83,Sanchez10},
was used to identify candidate states to explore.
The formal mathematics of the spin-state problem is identical
to that of the alloy problem.  Each site $i$ is given a 
parameter $\sigma_i$, where $\sigma_i = 1$ for spin up (species
A in the alloy problem) and $\sigma_i = -1$ for spin down
(species B in the alloy problem).

The total energy $U$ of a configuration 
$\{\sigma_i\}$
is written as
\begin{equation}
U (\{\sigma_i\}) = \sum_\alpha J_{\alpha} \xi_{\alpha},
\end{equation}
where $\alpha$ are the symmetry distinct geometric clusters, 
$J_{\alpha}$ the effective spin interaction parameter for 
cluster $\alpha$, and $\xi_{\alpha}$ the average spin
product over all symmetry equivalent occurrences of this cluster\cite{foot1}.

 In practice, one calculates individual $U$ for various configurations,
determines the values of $J_{\alpha}$ according to some fitting
procedure, and then uses these values to estimate the energies of any
configuration, including those that are not part of the fit.  Errors in the
determination of $J_{\alpha}$ due to the finite set of included energies
lead to errors in the predictions.
While any method for determining $J_{\alpha}$ should eventually converge to the same result
given a sufficiently large number of input configurations,
in this work, we employ the cluster expansion method formulated by Cockayne
and van de Walle\cite{Cockayne10}.  This method uses DFT results to fit a 
number of $J_{\alpha}$ parameters that is much {\em larger} than the number of
results.  The mathematical problem of overfitting the data (that is,
the non-uniqueness of the solution) is controlled by using a
Bayesian prior to constrain the magnitudes of the $J_{\alpha}$.
This method has additional advantages of exact fitting of all input energies
and self-consistent error estimates for all predictions.

 All results were calculated within an identical 80-atom, 32 Mn cell.
The cluster interactions that can be determined are limited to those
contained within one unit cell.  The model is valid for predictions of
other configurations within the same cell, but can not be applied to
larger supercells; thus if the ground state magnetic state has a larger
periodicity than the crystallographic one, it will be missed.  

There are $2^{32}$ collinear magnetic states for the 32 Mn atoms
in the \mn2o3 unit cell.  Time reversal and crystallographic symmetry
reduce the number of symmetry-independent configurations to about
3$\times$ 10$^8$, which, according to a one-to-one correspondence
with the number of cluster terms\cite{Connolly83}, yield about 
3$\times$ 10$^8$ unknown cluster terms.
Solving linear sets of equations with order 10$^8$ unknowns is not
computationally feasible.  To simplify the problem, we truncated the
interaction terms at fourth order (time reversal symmetry 
forbids linear or cubic terms in the magnetic interactions).  
There is 1 constant term, 73 independent pair cluster term
and 4632 four-body cluster terms in our model.
For sets of $n$ DFT total energy results, we solved $n$ equations
in 4706 unknowns, using a Bayesian prior to weight the parameters
and standard singular value techniques for solving underdetermined
sets of linear equations.

 The prior that we used for the pair terms was
\begin{equation}
P = \prod_{ij} {\rm exp} (-J_{ij}^2/(2 w_{ij}^2)),
\end{equation}
with 
\begin{equation}
w(i,j) = A \bigl( {{d_0}\over{d_{ij}}} \bigr)^2,
\end{equation}
where $d_{ij}$ is the distance between the Mn at site $i$ and $j$ and 
$d_0 \approx 3.3$~\AA$ = \sqrt{2} a_0/4$ is the
approximate nearest-neighbor Mn-Mn distance and $A$ is an unknown
constant.
The prior for the four-body terms was of similar form, with
\begin{equation}
w(i,j,k,l) = A    
\bigl( {{d_0}\over{d_{ij}}} \bigr)^2
\bigl( {{d_0}\over{d_{ik}}} \bigr)^2
\bigl( {{d_0}\over{d_{il}}} \bigr)^2
\bigl( {{d_0}\over{d_{jk}}} \bigr)^2
\bigl( {{d_0}\over{d_{jl}}} \bigr)^2
\bigl( {{d_0}\over{d_{kl}}} \bigr)^2
\end{equation}
The physical motivation behind the form of this interaction was to
weaken cluster terms involving Mn ions that are farther apart from
each other.  Although superexchange spin interactions 
are short range, we expect that strain coupling effects may mediate
longer range interaction; thus the $(d_{ij})^{-2}$
form for our relative interaction terms in the prior.
The value of $A$ was determined self-consistently by the 
leave-one-out cross validation method\cite{Cockayne10}. This
value was scaled such that the root mean square 
error in the predicted energies equalled the root mean square 
of the predicted errors\cite{Cockayne10}.

 We studied the magnetic states in an iterative manner.  After calculating an
initial set of energies versus magnetic orderings for a few simple
configurations, additional structures were investigated, with, in
rotating turns, (1) the minimum predicted energy among untested structures,
(2) the maximum predicted energy among untested structures, and
(3) the maximum predicted uncertainty in energy.
The parameters $J_{\alpha}$  were recalculated after each step, and then used
to predict the energies and energy uncertainties for {\em all} order
3$\times$ 10$^8$ symmetry independent configurations.
The model was refined iteratively until there were no more predicted states within
two standard deviations of uncertainty of the tenth lowest-energy state found,
at which point it was concluded that the collinear ground state was probably found.  
76 structures in all were calculated. 

\section*{Experimental Methods}


The \mn2o3 powder sample was prepared by heating 
MnCO$_3$ (analytical reagent) at 800 $^{\circ}$C in air for 12 h, which
was the highest temperature to yield phase-pure \mn2o3  devoid of 
Mn$_3$O$_4$ traces.   The sample was characterized using X-ray powder diffraction
in an instrument equipped with an incident-beam monochromator (Cu 
K$_{\alpha 1}$ radiation) and a position sensitive detector.  The heating temperature
and time were selected to minimize the width of the 222 peak, which
remains non-split in the orthorhombic phase and, therefore, reflects the
size of the coherently scattering domains in the sample.    (Phase-pure
\mn2o3 can be obtained by heating MnCO$_3$ at temperatures between 
600 $^{\circ}$C
and 800 $^{\circ}$C,
but lower temperatures produced considerably broader peaks).
No changes in the peak widths were observed after the second heating at
800 $^{\circ}$C for 12 h.  

Neutron powder diffraction measurements were performed
using both the time-of-flight HIPD diffractometer at the Lujan Center
of the Los Alamos National Laboratory and the BT-1 constant-wavelength
(Cu 311 monochromator, $\lambda = 1.5405$~\AA , 
15' collimation) diffractometer at the
NIST National Center for Neutron Research.   For these measurements,
the \mn2o3 powder was loaded in vanadium cans.  In each experiment the
data were collected at a series of temperatures (HIPD: 300 K, 200 K,
150 K, 100 K, 60 K, 40 K, 5 K and BT-1: 300 K, 100 K, 40 K, 10 K, 2 K).
Rietveld refinements of the nuclear and magnetic structures were
performed using GSAS\cite{GSAS}.    

The magnetic-structure models were selected
according to representational analyses performed by SARAh\cite{Sarah}; likewise, SARAh
was used for symmetry-constrained refinements in GSAS.   First, the
magnetic basis-vector coefficients were refined using a Reverse Monte
Carlo (RMC) algorithm implemented in SARAh with the magnitudes of all
the magnetic moments constrained to be equal.  The best-fit model was
further  refined in GSAS (i.e. using least squares minimization) by
keeping the basis-vector coefficients fixed but allowing for distinct
ordered magnetic moments on inequivalent Mn sites.    The HIPD and BT-1
data produced consistent structural parameters.  

 \section*{Computational Results}

 The \mn2o3 bixbyite structure was first investigated with
ferromagnetic ordering.  Relaxation under cubic \ia3 symmetry,
yielded $a_0$ = 9.409~\AA, and the structure shown in 
\tab{cubfer}.
DFT phonon results of this cubic structure show an extremely strong 
double instability ($\nu = 510~i~{\rm cm}^{-1}$),
associated with Jahn-Teller distortions of the oxygen octahedra
centered on the Mn(1) sites.    Full relaxation of the bixbyite
structure perturbed by either mode in the instability doublet, 
or any linear combination of the two, leads to an orthorhombic minimum 
energy state with $Pbca$ symmetry, explaining the experimental
cubic-orthorhombic transition.
In fact, we find that all such combinations relax to the {\em same}
ferromagnetic ground state structure, differing only with
respect to (1) which of the original cubic axis becomes the short axis
of the orthorhombic unit cell and (2) possible translations of the origin
by (1/2,1/2,1/2).   Each Mn(1) and Mn(2) in the orthorhombic structure
has four short Mn-O distances and two long Mn-O distances.
The topology of the orthorhombic structure is uniquely defined by
specifying which Mn(1)-O  and Mn(2)-O distances are long.
In this work, we arbitrarily choose the setting where the Mn(1) at 
(0,0,0) has its far O neighbors at approximately $\pm(0.139,0.149,-0.093)$, 
and the Mn(2) at (1/2,1/2,1/2) has its far O neighbors at 
approximately $\pm(0.656,0.412,0.641)$.  The ground state ferromagnetic
structure is shown in the left-hand side of \tab{orthofer}.

\begin{table} 
\caption{DFT coordinates for ferromagnetic \mn2o3 in cubic \ia3
phase. Lattice constant $a_0$ = 9.4090~\AA. }
\begin{tabular}{ccccc}
  \hline
Species & Site & $x$ & $y$ & $z$ \\
Mn(1) & 8($a$) &  0 & 0  & 0 \\
Mn(2) & 24($b$) & 0 & 1/4  & 0.2848  \\
O(1) & 24($b$) & 0.4162  & 0.1286  & 0.3555 \\
O(2) & 24($b$) & 0.3714  & 0.1445  & 0.0838 \\
\hline
\label{tab:cubfer}
\end{tabular} \end{table}

\begin{table}
\caption{Calculated DFT crystal structures for orthorhombic ferromagnetic
and ground state antiferromagnetic structures.  The similarity of $a$ and
$b$ for the orthorhombic ferromagnetic structure is coincidental.}
\begin{tabular}{cccccccc}
\hline
  &     &    &  FM  &    &    &  AFM ground &  \\
$a$ &    &     & 9.4417     &    &    &  9.4024         &  \\
$b$ &    &     & 9.4417      &    &    & 9.4435             &  \\
$c$ &    &     & 9.4096      &    &    & 9.3668             &   \\
Atom & Site  & x & y & z  &  x   &   y   &    z \\
Mn(1) & 4(a) &  0   &  0  &  0  &  0   &   0  &  0 \\
Mn(2) & 4(b) & 1/2  & 1/2  & 1/2  &  1/2  &  1/2  &  1/2  \\
Mn(3) & 8(c) &  0.2563 & 0.2854 & -0.0070 &  0.2602 & 0.2848 & -0.0102 \\
Mn(4) & 8(c) &  0.2864 & -0.0010 & 0.2458 &  0.2857 & -0.0034 & 0.2450 \\
Mn(5) & 8(c) &  0.0079 & 0.2478 & 0.2845 &  0.0136 & 0.2457 & 0.2818 \\
O(1) & 8(c) &  0.4215 & 0.1259 & 0.3510 &  0.4256 & 0.1233 &  0.3502 \\
O(2) & 8(c) &  0.1358 & 0.3523 & 0.4103 &  0.1409 & 0.3506 &  0.4050 \\
O(3) & 8(c) &  0.3576 & 0.4180 & 0.1223 &  0.3568 & 0.4187 &  0.1196 \\
O(4) & 8(c) &  0.0855 & 0.3721 & 0.1414 &  0.0845 & 0.3738 &  0.1395 \\
O(5) & 8(c) &  0.3791 & 0.1460 & 0.0794 &  0.3815 & 0.1478 &  0.0798 \\
O(6) & 8(c) &  0.1524 & 0.0859 & 0.3648 &  0.1567 & 0.0873 &  0.3608 \\
\hline
\label{tab:orthofer}
\end{tabular} \end{table}

\begin{table}
\caption{Ground state DFT magnetic state and lowest energy noncollinear 
magnetic state found in a Heisenberg model based on pair interactions
determined by fits to DFT results.  Units are ${\mu}_B.$  Only the
relative directions of the magnetic moments are determined; $m_{\parallel}$
is the magnetism along the only axis for the collinear DFT ground state and 
principal axis of the noncollinear Heisenberg model ground state; $m_{\perp}$
is the magnetism along a second axis of the Heisenberg model ground state.
Mn positions as in~\tab{orthofer}; spin moments for other Mn are 
related by applying the factors shown in~\tab{theomag2}.}
\begin{tabular}{ccccccccc}
\hline
      & DFT     &       & Heisenberg   &           &   \\
Atom  & $m_{\parallel}$ &       & $m_{\parallel}$  & $m_{\perp}$  & $\phi$ \\
Mn(1) & -3.6            &       & -3.6             &  0.0         &    0   \\
Mn(2) & -3.6    &       & -3.6   &  0.0   &    0   \\
Mn(3) &  3.6    &       &  3.0   & -1.9   &   32   \\
Mn(4) &  3.6    &       &  3.2   &  1.6   &   27   \\
Mn(5) & -3.6    &       & -3.6   & -0.1   &    1   \\
\hline
\end{tabular}
\label{tab:theomag}
\end{table}

\begin{table}
\caption{
Effects of $Pbca$ symmetry generators on the magnetization components 
of the theoretical \mn2o3 magnetic structures listed in \tab{theomag}.}
\begin{tabular}{ccc}
\hline
Generator        &   $m_{\parallel}$   &  $m_{\perp}$  \\
(x,1/2-y,1/2+z)  &  -1 &  -1 \\
(1/2+x,y,1/2-z)  &  -1 &  -1 \\
(1/2-x,1/2+y,z)  &  +1 & -1 \\
\hline
\end{tabular}
\label{tab:theomag2}
\end{table}

The orthorhombic crystallographic structure of the ferromagnetic phase
was used as the starting part for relaxation of each collinear 
spin combination $\{\sigma_i\}$ studied.  
The state that was ultimately picked as the ground state was the 36th studied.
In total, 76 states were investigated before the termination criterion was
reached:  no new structures with predicted energies within two
standard deviations of the tenth-lowest-energy structure found.

\begin{figure}[h]
\includegraphics[width=75mm]{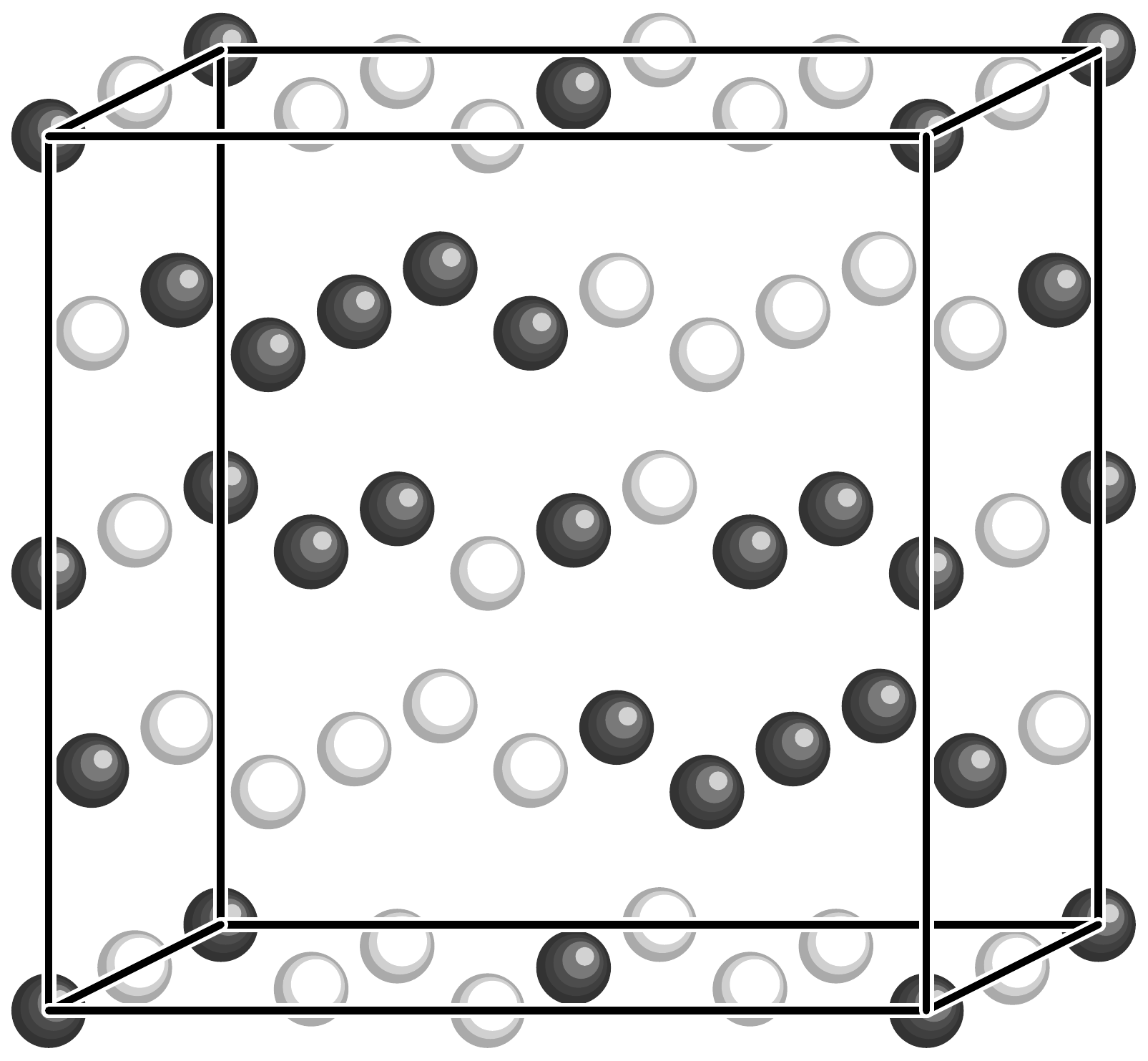}
\caption{Lowest-energy collinear magnetic structure found
computationally; also collinear model that gives best fit 
$(\chi^2 = 2.54)$
to
experimental results at 2 K.  Only Mn atoms shown; dark spheres represent
spin ``up" and light spheres spin ``down".}
\label{fig:afm}
\end{figure}

The calculated ground state collinear magnetic structure is shown in \fig{afm}
and listed in \tab{theomag}.  The spin moments are obtained by taking the
difference in the number of spin up and spin down electrons, 
integrated within spheres of radius 1.24~\AA~centered on each Mn, as
calculated using VASP.
The magnitudes are reasonable for high-spin Mn$^{3+}$ ions, but 
are not precisely comparable with experiment because of the
artificial partition of the cell into spherical volumes.

\begin{figure}[h]
\includegraphics[width=75mm]{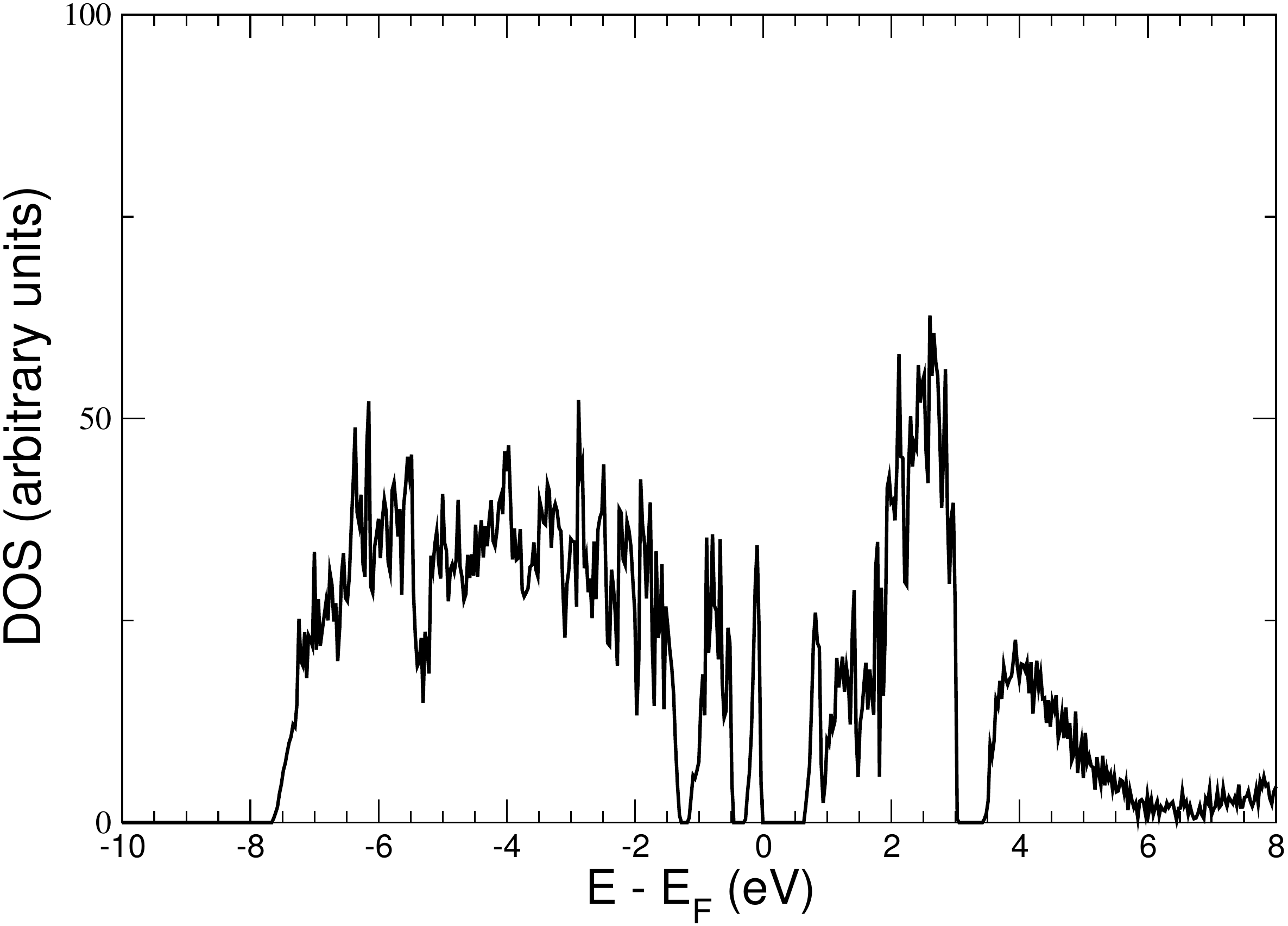}
\caption{Calculated density of states (DOS) for \mn2o3 in
ground state collinear magnetic structure.}
\label{fig:dossy}
\end{figure}

 We next investigated possible noncollinear magnetism using the
parameters found in the fit to the computational results.  The
obvious way to extend the model is to use the same parameters for
the collinear approach, but to vectorize the spins in a Heisenberg
model approach.  That is, instead of using Ising-like interaction 
terms $J_{ij} \sigma_i \sigma_j$, one uses Heisenberg-like interaction
of spin ``vectors" according to the principle axes of ``up" spin:
$J_{ij} \vec{\sigma}_i \cdot \vec{\sigma}_j$, with exactly the
same set of $J_{ij}$.  For simplicity, we only included pair interactions
in this approach.   Simulated annealing was used to find the lowest
energy state of the Heisenberg model.  Random initial spin configurations
always converged to an equivalent noncollinear ground state
(right side of \tab{theomag}), with a dominant spin axis direction 
$m_{\parallel}$, and all secondary spin components along the same
orthogonal axis $m_{\perp}$.  

 The band structure for the lowest-energy collinear magnetic state 
found is shown in \fig{dossy}.
The structure has some similarities to that of 
Franchini {\it et al.}\cite{Franchini07}, using
the PBE0 approximation, including a band gap at the Fermi level (0.6 eV
in our case).  Our results differ from theirs in that our 
strict antiferromagnetism leads to a band structure without a
distinction between majority and minority spins.  Furthermore, we find a remarkable 
signature of the combination of octahedral crystal field splitting and effects
of Jahn-Teller distortion in the nature of the highest occupied levels:
they are split off from the other occupied 3$d$ states. The highest
valence band is an isolated band associated with 3$d$ electrons of the
Mn(1) and Mn(2).  A second split band at around $E_F -0.8$ eV is associated
with Mn(3), Mn(4), and Mn(5) 3$d$ electrons.

\section*{Experimental Results}

\begin{figure}[]
\includegraphics[width=75mm]{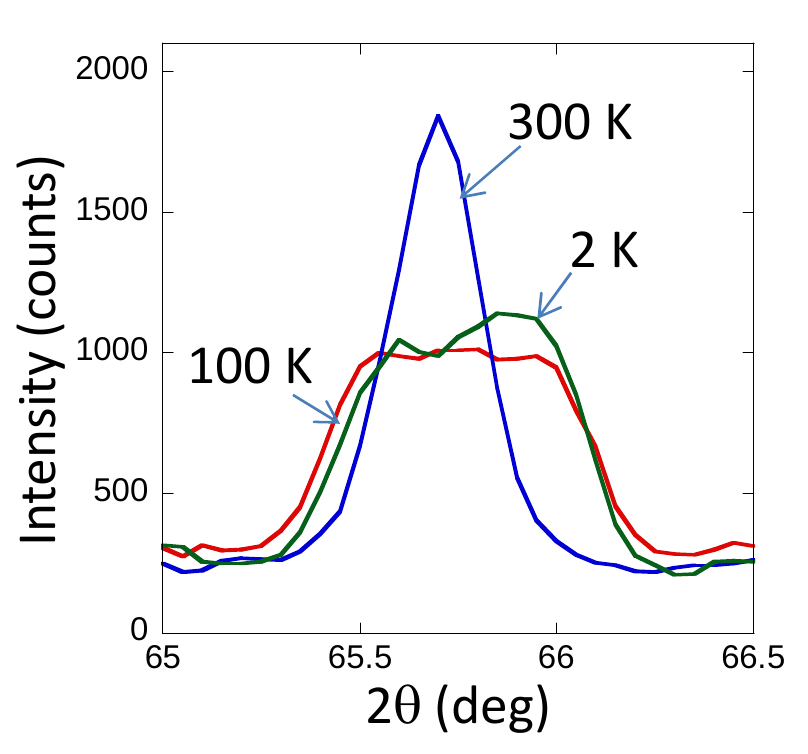}
\caption{A trace of the 622 cubic reflection at three different
temperatures.  This reflection appears as a single peak at 300 K but
exhibits pronounced splitting at sub-ambient temperatures.   Similar
trends are observed for other reflections.}
\label{fig:igor1}
\end{figure}

\begin{figure}[]
\includegraphics[width=75mm]{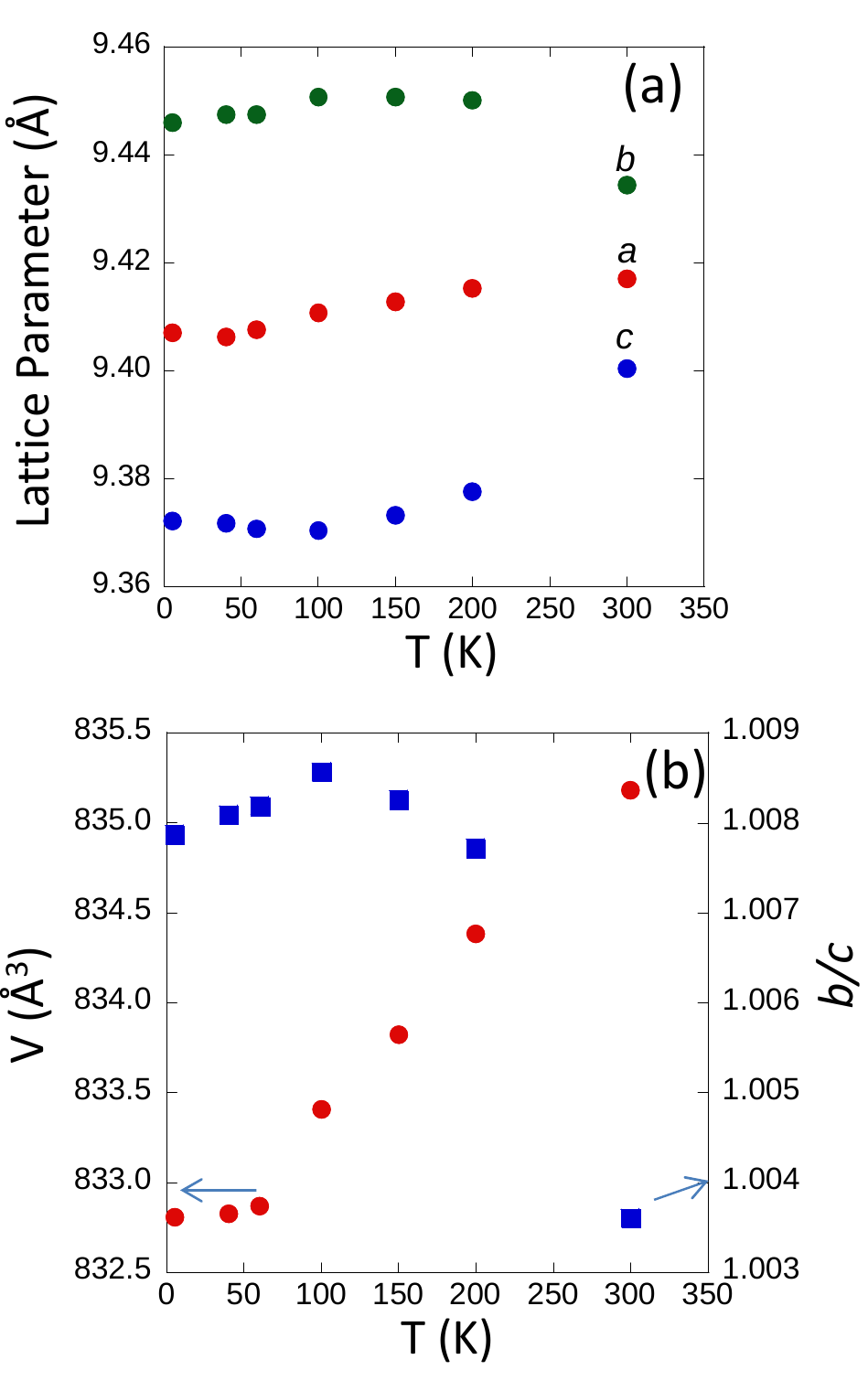}
\caption{Temperature dependence of the (a) orthorhombic lattice parameters
and (b) unit-cell volume and $b/c$ ratio which characterizes the magnitude
of the orthorhombic distortion. The behavior of the orthorhombic
distortion changes across the magnetic transition at approximately 80 K.
The error bars are within the size of the symbols.}
\label{fig:igor2}
\end{figure}

\begin{figure}[]
\includegraphics[width=75mm]{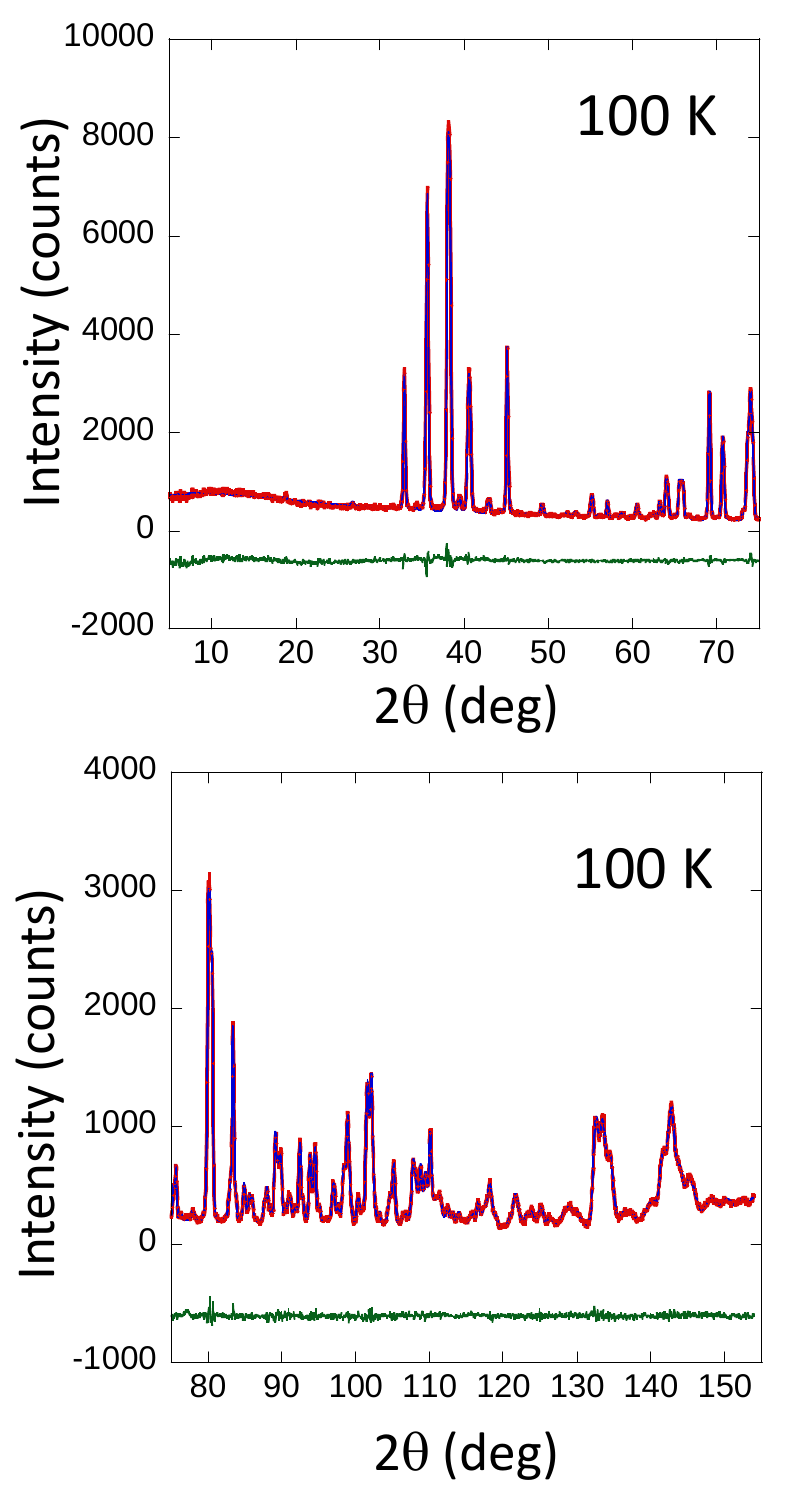}
\caption{Experimental (red/dots) and calculated (blue/line) neutron diffraction
profiles (BT-1) for \mn2o3 at 100 K.   The residual is indicated
below (green line).  The agreement factors are $\chi^2 = 1.16$ and
$R_{wp} =4.79 \%$.}
\label{fig:igor3}
\end{figure}

Room-temperature diffraction patterns (\fig{igor1}) exhibit no clear reflection
splitting indicative of an orthorhombic distortion.  However, the peaks
are broad and the cubic \ia3  model fits poorly yielding abnormally large
atomic displacements parameters $U_{iso}$ for the oxygen atoms, which suggests
that the structure is distorted.   The data (not shown) can be fitted
satisfactorily using the orthorhombic $Pbca$ structure reported in the
literature with sensible $U_{iso}$ values.  The refined lattice distortion 
at
300 K differed considerably between the HIPD data $(b/c = 1.0007)$ and BT-1 
data $(b/c = 1.0036)$,
presumably because of the close proximity of the phase transition
which leads to relatively large changes in the lattice distortion even
for small temperature differences. The distortion increases rapidly on
cooling to $b/c = 1.0085$ at 100 K and the reflection splitting becomes evident
(\fig{igor1}).  For these temperatures, the lattice parameters refined using
the HIPD and BT-1 data were in good agreement.  Temperature dependencies
of the lattice parameters, unit cell volume, and $b/c$ data
are summarized in~\fig{igor2}. \fig{igor3} displays the experimental and
calculated diffraction profiles for T=100 K, while~\tab{igor1} summarizes
the results of the nuclear-structure refinements at 100 K and 2 K.
The MnO$_6$ octahedra exhibit strong Jahn-Teller distortions already at
300 K (\tab{igor2}).  

\begin{table*}[t]
\centering
\caption{
 Parameters of the nuclear structures of paramagnetic (T=100 K)
and antiferromagnetic (T=2 K) \mn2o3 obtained by Rietveld refinements
using neutron powder diffraction data (BT-1). In both cases, the space
group is $Pbca$ ($\#$61).  The model assumed isotropic atomic displacement
parameters ($U_{iso}$), which were constrained according to the atom type
(i.e. Mn or O).  The refined values were  $U_{iso}$(Mn) =  0.0031(2) \AA$^2$
and $U_{iso}$(O) = 0.0049(1)~\AA$^2$ at 100 K and $U_{iso}$(Mn) = 0.0021(3)~\AA$^2$
and $U_{iso}$(O) = 0.0040(1)~\AA$^2$ (O) at 2 K.  Numbers in parentheses refer to
one standard deviation as calculated in GSAS.}
\begin{tabular}{cccccccc}
\hline
  &     &    &  100 K ($\chi^2 = 1.16$)  &    &    &  2 K ($\chi^2 = 2.54$) &  \\
\hline
a &    &     & 9.4104(1)     &    &    &  9.4078(1)         &  \\
b &    &     & 9.4509(1)      &    &    & 9.4488(1)             &  \\
c &    &     & 9.3706(1)      &    &    & 9.3739(1)             &   \\
\hline
Atom & Site  & x & y & z  &  x   &   y   &    z \\
Mn(1) & 4(a) &  0   &  0  &  0  &  0   &   0  &  0 \\
Mn(2) & 4(b) & 1/2  & 1/2  & 1/2  &  1/2  &  1/2  &  1/2  \\
Mn(3) & 8(c) &  0.2598(5) & 0.2854(3) & -0.0105(5) &  0.268(1) & 0.280(1) & -0.011(1) \\
Mn(4) & 8(c) &  0.2867(3) & 0.0017(8) &  0.2436(4) &  0.2880(9) & 0.009(2) & 0.246(1) \\
Mn(5) & 8(c) &  0.0127(5) & 0.2456(6) &  0.2827(3) &  0.012(1) & 0.237(1) & 0.2833(9) \\
O(1) & 8(c) &  0.4253(3)  & 0.1241(4) &  0.3511(4) &  0.4276(8) & 0.1240(9) &  0.3502(8) \\
O(2) & 8(c) &  0.1399(3)  & 0.3509(4) &  0.4068(2) &  0.1395(8) & 0.3513(9) &  0.4074(7) \\
O(3) & 8(c) &  0.3579(3)  & 0.4175(4) &  0.1218(3) &  0.3586(9) & 0.4169(9) &  0.1209(7) \\
O(4) & 8(c) &  0.0838(3)  & 0.3725(4) &  0.1402(3) &  0.0738(7) & 0.3712(9) &  0.1385(9) \\
O(5) & 8(c) &  0.3797(3)  & 0.1477(4) &  0.0789(3) &  0.3821(8) & 0.1509(9) &  0.0803(7) \\
O(6) & 8(c) &  0.1554(3)  & 0.0882(3) &  0.3609(4) &  0.1563(8) & 0.0882(9) &  0.3586(8) \\
\hline
\end{tabular}
\label{tab:igor1}
\end{table*}

\begin{table}
\caption{
Mn-O distances at 300 K and 100 K (in~\AA).
Numbers in parentheses refer to one standard deviation as calculated in GSAS.}
\begin{tabular}{cccccc}
\hline
     & 300 K   &      &    & 100 K  &    \\
Atom &   &      &    &        &    \\
Mn(1) & 2.03(2) ($\times$2) &          &   & 1.949(3) ($\times$2)  &        \\
      & 2.02(1) ($\times$2) &          &   & 1.944(3) ($\times$2)  &        \\
      & 1.97(2) ($\times$2) &          &   & 2.129(3) ($\times$2)  &        \\
Mn(2) & 2.04(2) ($\times$2) &          &   & 1.953(3) ($\times$2)  &        \\
      & 1.90(1) ($\times$2) &          &   & 2.117(3) ($\times$2)  &        \\
      & 2.04(2) ($\times$2) &          &   & 1.923(3) ($\times$2)  &        \\
Mn(3) & 2.17(2)             & 1.88(1)  &   & 2.200(5)              & 1.879(5) \\
      & 2.03(2)             & 2.31(2)  &   & 1.987(5)              & 2.327(5) \\
      & 1.92(1)             & 2.00(2)  &   & 1.915(5)              & 1.961(5) \\
Mn(4) & 2.00(2)             & 2.20(2)  &   & 2.033(6)              & 2.181(7) \\
      & 1.90(1)             & 1.96(2)  &   & 1.934(5)              & 1.959(6) \\
      & 2.35(2)             & 1.88(1)  &   & 2.269(7)              & 1.859(5) \\
Mn(5) & 1.90(2)             & 1.96(2)  &   & 1.889(5)              & 1.943(5) \\
      & 2.26(2)             & 1.91(2)  &   & 2.358(6)              & 1.916(5) \\
      & 1.98(2)             & 2.18(2)  &   & 2.026(5)              & 2.134(5) \\
\hline
\end{tabular}
\label{tab:igor2}
\end{table}

\begin{figure}[]
\includegraphics[width=65mm]{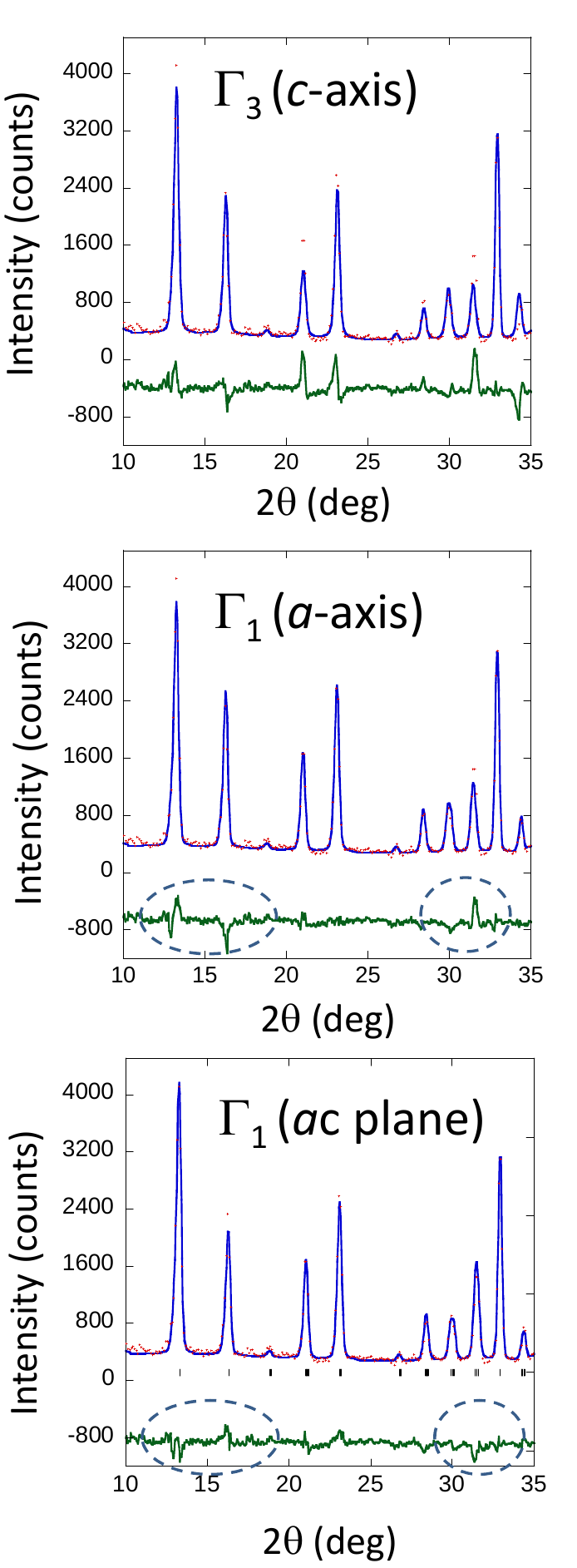}
\caption{A low-angle portion of the neutron diffraction pattern collected
at 2 K showing experimental (red/dots) and calculated (blue/line) neutron
diffraction profiles.  Note that all the reflections with $2 \theta 
\leq 32^{\circ}$ were absent at 100 K (\fig{igor3}); these reflections, all indexable
according to the primitive nuclear-structure unit cell, originate from
magnetic ordering.   The calculated profiles correspond to (top) the $\Gamma_3$
model by Regulski et al. with magnetic moments aligned with the c-axis,
(middle) the $\Gamma_1$ model with magnetic moments collinear with the a-axis,
and (bottom) the $\Gamma_1$ model with non-collinear magnetic moments residing
in the ac plane.   The $\Gamma_1$ models provide a superior fit relative to the
$\Gamma_3$ model.   A non-collinear alignment of magnetic moments significantly
improves the fit for several magnetic reflections (outlines using a
dashed line).}
\label{fig:igor4}
\end{figure}

Below 100 K, a series of strong reflections appears at
lower angles, which signifies magnetic ordering (\fig{igor4}).   
The $b/c$ ratio
decreases slightly below the magnetic transition 
(\fig{igor2}(b))
to 1.0079 at 2 K.  The patterns remain qualitatively unchanged from 60 K 
down to 2 K.   The magnetic reflections can be accounted for by a propagation
vector k=0.   All eight symmetry elements of the space group $Pbca$ leave
this propagation vector invariant.     Group-theory analysis yields eight
irreducible representations (IR) ($\Gamma_i$, $i$=1, 8), each having an order
of one.   Only four of these IRs ($\Gamma_1$,
$\Gamma_3$, $\Gamma_5$, and $\Gamma_7$),
are common to all five of the inequivalent Mn sites.  

 Regulski {\it et al.}\cite{Regulski04} presented their collinear magnetic-ordering model in the
 form of schematic drawings for each Mn sublattice; no symmetry analysis
was performed.   According to our representational analysis, the
 ordering types for their sublattices 1 (combined Mn 4(a) and 4(b) sites),
 2, and 4 belong to the $\Gamma_3$ representation.   However, the ordering
 on their sublattice 3 is incompatible with any of the IRs; that is, the
 model, at least as presented, is not compatible with orthorhombic
crystallographic symmetry  (possibly there is a drawing error).
 The spin arrangement for their sublattice 3 can be made compatible with
 the structural symmetry by flipping the directions of two spins.
 The resulting ordering patterns belong to either the $\Gamma_3$ or 
$\Gamma_6$ representations, depending on which two spins are flipped.
 We fitted both models ({\it i.e.}  
$\Gamma_3$  and $\Gamma_3 + \Gamma_6$) to our data.
 A collinear pattern with magnetic moments directed along one of the
 orthorhombic axes was assumed.   The $\Gamma_3$ model with all the moments
 parallel to the $c$-axis produced a superior fit of quality comparable to
 that reported by Regulski {\it et al.}   However, examination of the misfit
 between the calculated and experimental profiles reveals significant
 discrepancies for several magnetic reflections that become split in the
 orthorhombic structure (\fig{igor4}(a)), which indicates that the intensity
 distribution among the split-peak components is incorrect.   Conceivably,
 these deficiencies were obscured by the insufficient resolution in the
 data used by Regulski {\it et al}.

As the literature model failed to describe the data, we considered
collinear models corresponding to other representations (i.e. $\Gamma_1$,
 $\Gamma_5$ and $\Gamma_7$).   The $\Gamma_1$ model with the magnetic moments directed along
the $a$-axis provided a satisfactory fit to the neutron data of quality
far superior to that obtained for any of the $\Gamma_3$ models (\fig{igor4}(b));
the collinear $\Gamma_1$ models with magnetic moments directed along the $b$ and
$c$-axes yielded considerably worse agreement factors.  Models generated
according to the $\Gamma_5$ and $\Gamma_7$ representations generated poor fits
and were discarded.   Refinements of the magnetic-moment magnitudes
$(m)$ independently for each Mn site significantly improved the fit.
The resulting $m$-values, which at 2 K range from approximately 
3 $\mu_B$  to 4 $\mu_B$,
are consistent with those expected for Mn$^{3+}$ ions.  
The collinear model most consistent with the experimental results is
shown in \tab{igor3}.  It is {\em identical} to that determined
computationally (\fig{afm}).

\begin{table}
\caption{
Magnetic-moment (m) components for the Mn atoms in the best-fit
$\Gamma_1$ models at 2 K and 40 K.  The atomic coordinates at 2 K are given
in~\tab{igor1}.
The magnetic-ordering model assumed $m_y$=0.  The refined magnitudes of $m$
(units of $\mu_B$) are indicated for the inequivalent Mn sites.
Magnetic moments for the remaining Mn sites are generated as
described in \tab{igor4}.
The refinements at 2 K and 40 K were performed independently with
the random starting models for the basis-vector mixing coefficients.
The angle between the $m_{tot}$ and the a-axis is defined as
$\phi$ ($^{\circ}$). Numbers in parentheses refer to
one standard deviation as calculated in GSAS.}
\begin{tabular}{ccccccccc}
\hline
              &       & 2 K   &           &        &       & 40 K  &           &       \\
Atom        & $m_x$ & $m_z$ & $m_{tot}$ & $\phi$ & $m_x$ & $m_z$ & $m_{tot}$ & $\phi$ \\
Mn(1) & -2.6  &  1.6  &  3.1(1)   &  32    &  -2.3 &  1.4  &  2.7(1)   &   32   \\
Mn(2) & -3.4  & -0.7  &  3.5(1)   &  12    &  -3.0 & -0.8  &  3.1(1)   &   15   \\
Mn(3) &  3.2  & -1.4  &  3.5(1)   &  23    &   3.0 & -0.3  &  3.0(1)   &    5   \\
Mn(4) &  3.0  &  1.3  &  3.3(1)   &  24    &   2.9 &  0.3  &  2.9(1)   &    6   \\
Mn(5) & -3.5  & -2.3  &  4.2(1)   &  34    &  -3.4 & -1.0  &  3.5(1)   &   16   \\
\hline
\end{tabular}
\label{tab:igor3}
\end{table}

\begin{table}
\caption{
Effects of generators of $Pbca$ symmetry on magnetic moments refined experimentally.}
\begin{tabular}{ccc}
\hline
Generator        &   $m_x$   &  $m_z$   \\
(x,1/2-y,1/2+z)  &  -1 &  -1 \\
(1/2+x,y,1/2-z)  &  -1 &  +1 \\
(1/2-x,1/2+y,z)  &  +1 & -1 \\
\hline
\end{tabular}
\label{tab:igor4}
\end{table}

We explored
potential deviations from the collinearity using the algorithms
implemented in SARAh.    A relatively large number of RMC cycles
(3,000 for 2 basis-vector mixing coefficients per site and 10,000
for 3 coefficients per site) were found necessary to locate a model
that provides an adequate fit (a goodness-of-fit $\chi^2 \le 3$) to the data.
Detailed refinements of the magnetic structure were performed at 40 K and
2 K.  Multiple refinements that start from randomly selected values of
the mixing coefficients were run to verify the robustness of the best-fit
structural model at 2 K.  

\begin{figure}[]
\includegraphics[width=75mm]{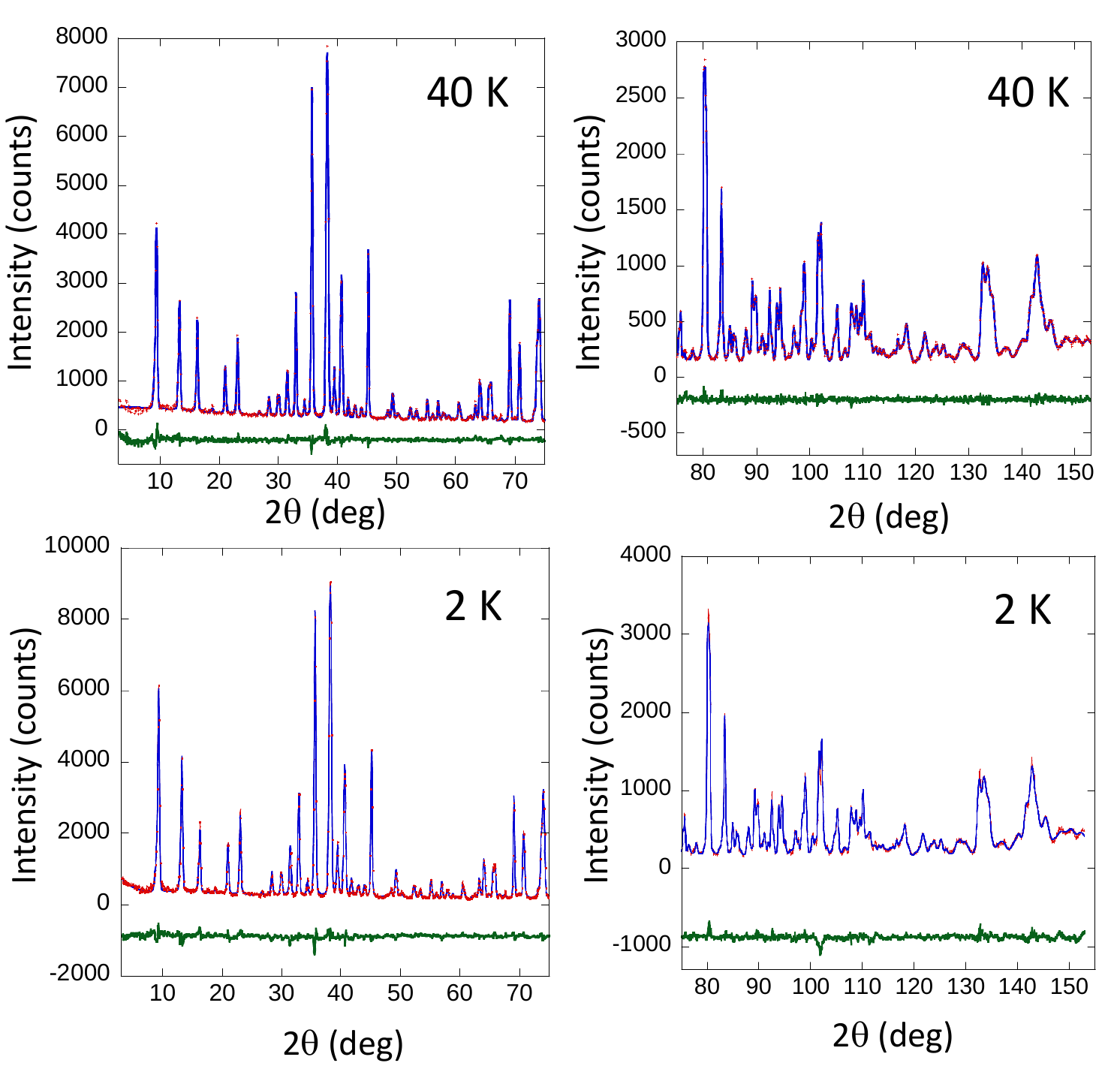}
\caption{Experimental (red/dots) and calculated (blue/line)
neutron diffraction profiles (BT-1) for \mn2o3 at 40 K (top) and 2 K
(bottom).   The calculated profiles correspond to the $\Gamma_1$ model with the
non-collinear array of magnetic moments in the ac plane.    The agreement
factors are $\chi^2 = 1.22$ and $R_{wp} = 5.21 \%$ (40 K) and 
$\chi^2=2.53$ and $R_{wp} = 6.89 \%$ (2 K).}
\label{fig:igor5}
\end{figure}

The best fit was obtained by restricting the
magnetic moments to the orthorhombic (010) plane.  The fit of the several magnetic
reflections was visibly improved in the non-collinear model (\fig{igor4}(c),
\fig{igor5}).  The best-fit $\Gamma_1$ models at both 40 K and 2 K feature similar
patterns of magnetic ordering (\tab{igor1}).  No additional improvement
was obtained by varying all three mixing coefficients per site after
20,000 RMC cycle; possibly, even this number of cycles was insufficient
to identify a global minimum.   In the best-fit non-collinear model,
the magnetic moments are aligned preferentially with the $a$-axis but
exhibit significant (up to 32$^{\circ}$) deviations from this direction;
the deviations from collinearity become especially pronounced at 2 K.

\section*{Discussion}

\begin{figure}[h]
\includegraphics[width=75mm]{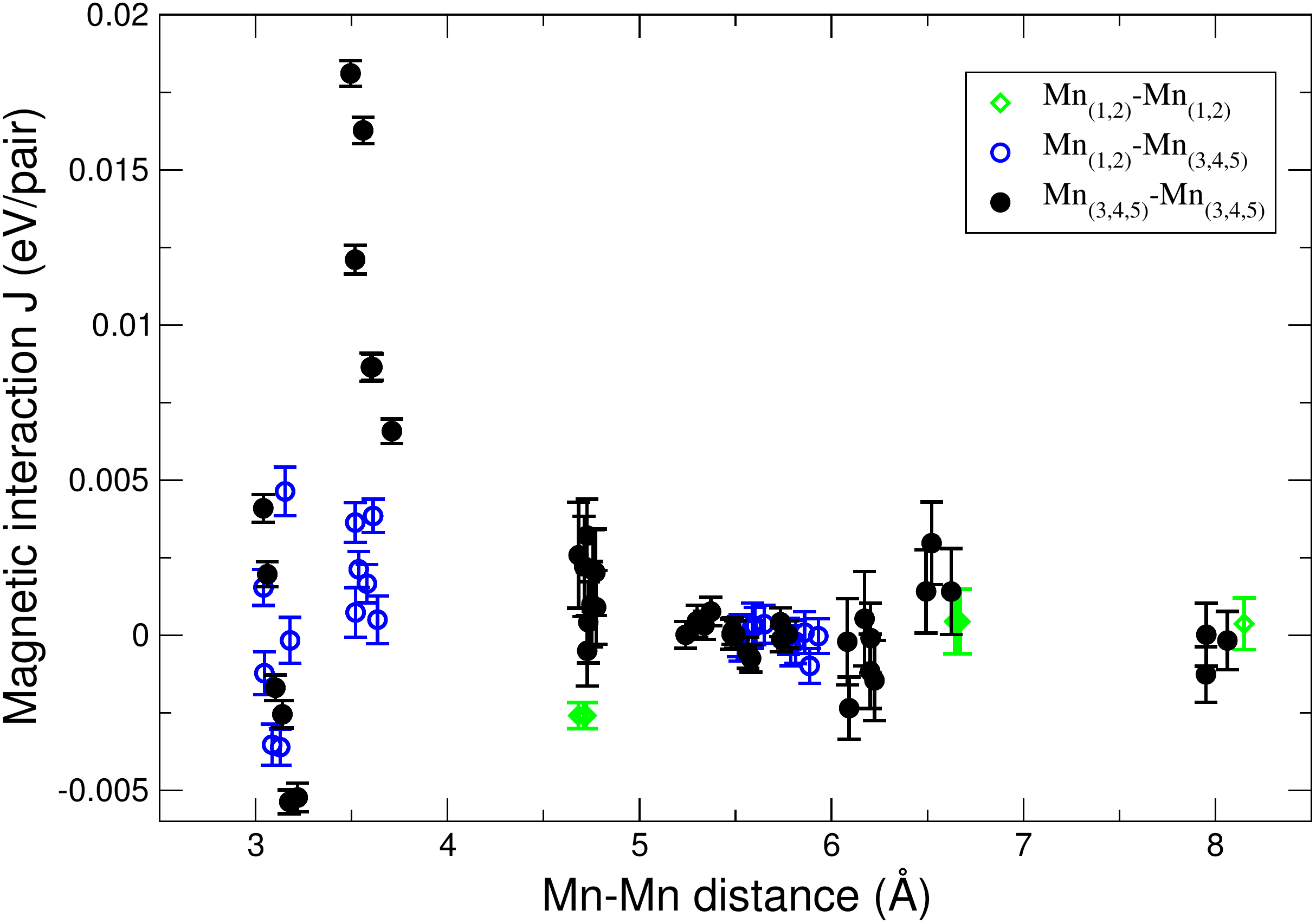}
\caption{Calculated magnetic interaction parameters between Mn
ions in \mn2o3.  Positive values for $J$ favor antiferromagnetic
alignment.  Mn subscripts refer to the different Mn sites in the 
orthorhombic phase.  Error bars indicate plus and minus one standard 
deviation of the parameter, based on cross-validation
calculations.}
\label{fig:magplot}
\end{figure}

 The theoretical lattice parameters of orthorhombic \mn2o3  change
significantly when the magnetic structure changes from 
ferromagnetic to the antiferromagnetic 
lowest-energy collinear state (\tab{orthofer}), demonstrating significant
spin-strain coupling.  The ground state lattice parameters are
within about $0.1\%$ of experiment, phenomenally good agreement that
demonstrates the accuracy of DFT calculations
using the PBEsol exchange correlational along with on-site $U$ and
$J$ parameters to treat $d$-electron correlations in Mn.

 The lowest DFT collinear state found
and the best low-temperature experimental fit to a collinear
model are identical, although they were achieved completely
independently, suggesting that the nature of the magnetism in
\mn2o3 is largely solved.
The secondary components of the  best noncollinear spin arrangements
of model and experiment (\tab{theomag} and \tab{igor3}) appear different
at first glance, but in fact, the two are related by the
approximate relationship
$m_{\perp} ({\rm model}) \approx  {\rm cos} (2 \pi x) m_{\perp} ({\rm expt.})$,
with $x$ the crystallographic positional coordinate.
The source of the discrepancy is not clear, but the agreement is
noteworthy given the approximations involved in the computational 
approach.  Note that the experimental results show clear preferences
for magnetic moments along particular {\em directions}; the
DFT approach neglected spin-orbit coupling, and thus the
effect of magnetic moment direction could not be studied.

\begin{figure}[h]
\includegraphics[width=75mm]{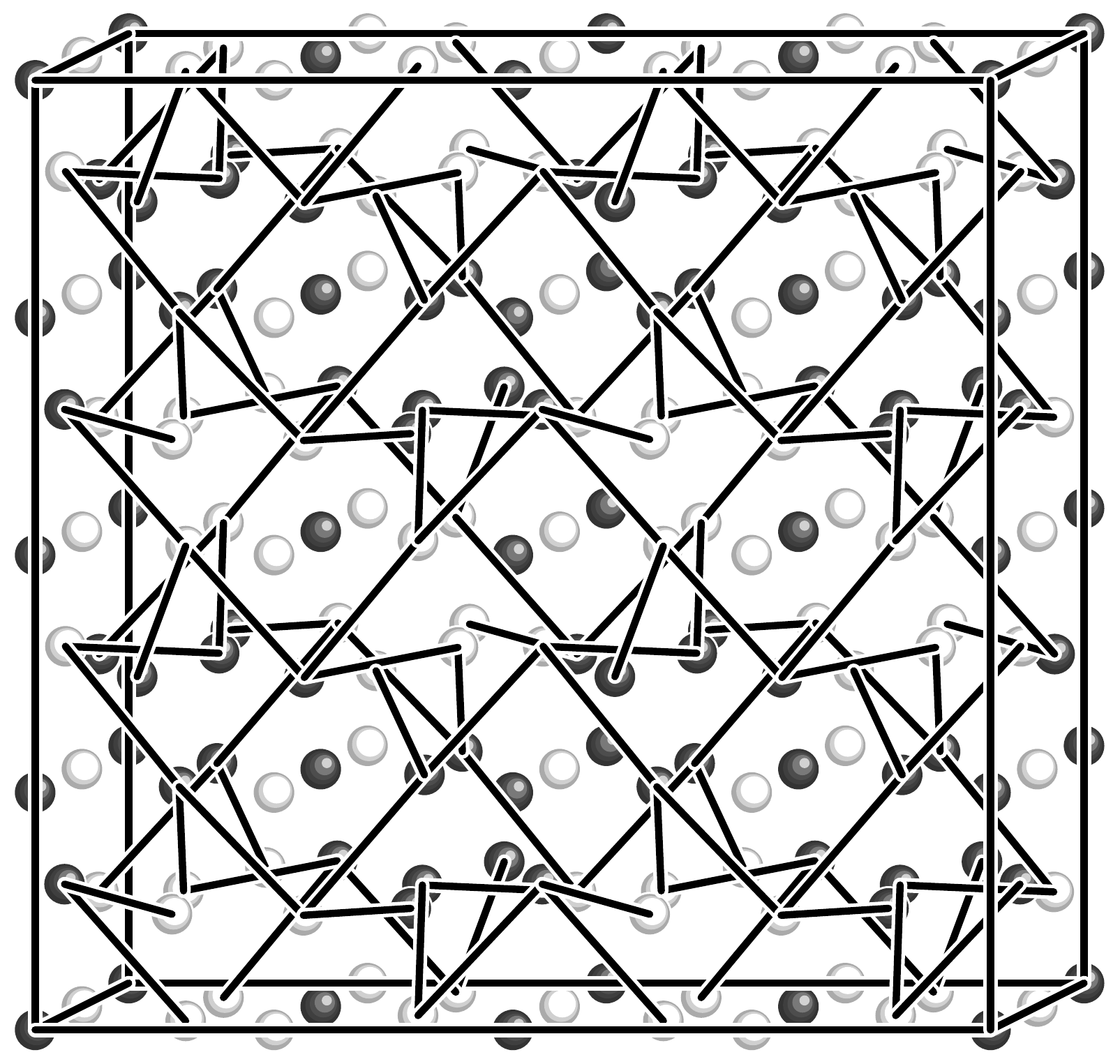}
\caption{Linkages between Mn in \mn2o3 limited to 
a subset of those with strong antiferromagnetic
interactions as determined in this work gives the lowest
energy collinear magnetic structure on the Mn(3), Mn(4), and
Mn(5) sites.}
\label{fig:satisfy}
\end{figure}

\begin{figure}[h]
\includegraphics[width=75mm]{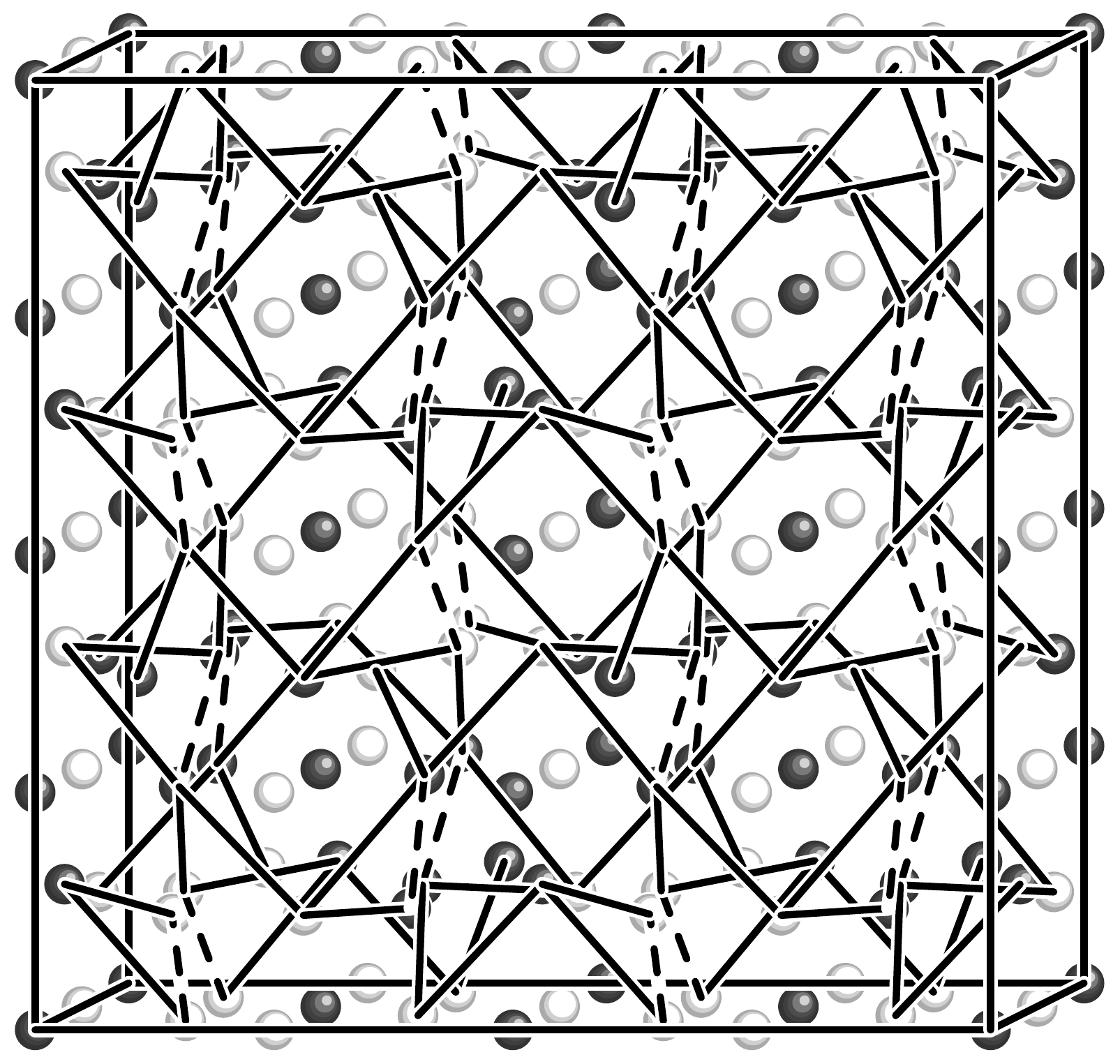}
\caption{Similar to \fig{satisfy}, with the Mn-Mn pairs
with the five strongest AFM interactions shown.  One
additional interaction, not shown in \fig{satisfy}, is
indicated by dotted lines here, and leads to frustrated triangles,
which presumably is the origin of the noncollinear ground state
magnetism.}
\label{fig:frustrate}
\end{figure}

 The pair magnetic parameters, as determined from the fit to
the DFT results, are shown as a function of Mn-Mn distance 
in \fig{magplot}.  The strongest terms are 
antiferromagnetic terms for exactly those Mn-Mn pairs that 
(1) are approximately 3.5~\AA~to 3.6~\AA~apart,
(2) share one close O atom bonded to each, and (3) 
have a Mn-O-Mn angle for these bonds less than 125$^{\circ}$.  
These Mn-O-Mn pairs involve two Mn(3), Mn(4), or Mn(5)  atoms and have very
unequal Mn-O distances.  Within the Mn-Mn pairs
the magnitude of the antiferromagnetic interaction
strictly decreases with increasing length of
the longer Mn-O distance.

 Two pair interactions of approximate magnitude 0.085 eV per pair
are nearly degenerate in energy and Mn-Mn distance, such that they
can not be distinguished in \fig{magplot}.
If the Mn-Mn links corresponding to the 3 strongest terms and
one of the 0.085 eV terms are drawn (\fig{satisfy}),
then the magnetic structure for the Mn(3), Mn(4), and
Mn(5) sites based on these interactions, all antiferromagnetic, 
is completely determined to be that of the
computational and experimental collinear ground state.
On the other hand, if links corresponding to both 0.085 eV terms are
drawn (\fig{frustrate}), then there are frustrated triangles where
antiferromagnetic interactions can not be satisfied for all bonds.
A triangle of vector spins with frustrated
antiferromagnetic interactions has a (possible degenerate) 
ground state with a noncollinear spin arrangement; we
believe this is the origin of the noncollinear antiferromagnetism
of \mn2o3. 

 We only looked at the ground state magnetic structure, but the
finite temperature magnetic correlations could also be investigated
using the DFT model, which would allow the nature of possible
AFM-AFM transitions in \mn2o3 (\onlinecite{Regulski04}) to be
determined.

\label{sec:discussion}

\section*{Conclusions}

First principles density functional theory DFT+U and cluster expansion model calculations,
along with independent experimental neutron diffraction structure analyses, were used to determine 
the low-temperature crystallographic and magnetic structure of bixbyite \mn2o3.  
Both approaches independently  gave nearly identical crystallographic and
magnetic structures, with identical antiferromagnetic ordering along a
principal magnetic axis and secondary ordering along a single orthogonal axis,
differing only by a phase factor in the modulation patterns.
The agreement between the two approaches suggests that the ground
magnetic state of \mn2o3 is largely solved.   

The computational methods
exploited a Bayesian approach that allows the number of parameters
in the cluster expansion model to exceed the number of input structure energies without 
sacrificing energy predictability.   The individual magnetic coupling parameters were determined,
showing that specific frustrated antiferromagnetic interactions determine the magnetic
structure.  The experimental approach benefited from optimized sample synthesis,
which produced crystallite sizes large enough to yield a clear splitting of peaks in the
neutron powder diffraction patterns, thereby enabling magnetic-structure refinements under
the correct orthorhombic symmetry.

The approaches used here should prove suitable for similar problems of magnetic ordering in
other complex oxides whose magnetic states are determined by a large and 
frustrated set of antiferromagnetic interactions.

\section*{Acknowledgements}

This work has benefited from the use of HIPD at the Lujan Center at Los Alamos                                                                                    
Neutron Science Center, funded by DOE Office of Basic Energy Sciences.  Los                                                                                       
Alamos National Laboratory is operated by Los Alamos National Security LLC                                                                                        
under DOE Contract DE-AC52-06NA25396.


\begin{thebibliography}{19}
\bibitem[{foo({\natexlab{a}})}]{foot2}
\bibinfo{note}{The Table in the Landoldt-B\"ornstein(\onlinecite{LB75}) has the
  $b$ and $c$ parameters transposed with respect to the original Geller and
  Espinosa paper(\onlinecite{Geller70}). Based on our results, we believe that
  the ordering in the original Geller and Espinosa work is correct for the
  standard Pbca space group setting.}
\bibitem[{dis()}]{disclaim}
\bibinfo{note}{Certain commercial software is identified in this paper to
  adequately describe the methodology used. Such identification does not imply
  recommendation or endorsement by the National Institute of Standards and
  Technology, nor does it imply that the software identified is necessarily the
  best available for the purpose.}
\bibitem[{foo({\natexlab{b}})}]{foot1}
\bibinfo{note}{For clarity, cluster multiplicities due to symmetry are
  neglected in the text equations, but were properly weighted in the analysis.}
\end{thebibliography}
\end{document}